\documentclass{article}

\usepackage{PRIMEarxiv}
\usepackage[utf8]{inputenc}
\usepackage[T1]{fontenc}
\usepackage{microtype}

\usepackage{cite}
\usepackage{amsmath,amssymb,amsfonts}
\usepackage{booktabs}
\usepackage{array}
\usepackage{multirow}
\usepackage{makecell}
\usepackage{graphicx}
\usepackage[export]{adjustbox}
\usepackage[usenames,dvipsnames]{xcolor}
\usepackage[normalem]{ulem}
\usepackage{pifont}
\usepackage{subfig}
\usepackage[hyphens]{url}
\usepackage{hyperref}

\newcommand{\fig}[1]{Fig.~\ref{#1}}

\newcommand{\sys}{\textsc{PhOSv1}}
\newcommand{\crbaseline}{\textsc{PhoenixOS}}

\newcommand{\blue}[1]{\textcolor{blue}{#1}}

\providecommand{\burl}[1]{\url{#1}}

\begin{document}

\title{\sys: Efficient Scaling of Model Sleeping on a Single Server for Autoscaling with GPU-Context Deduplication}

\author{
Tianle Sun \\
Shanghai Jiao Tong University \\
China
\And
Xingda Wei \\
Shanghai Jiao Tong University \\
China
\And
Rong Chen \\
Shanghai Jiao Tong University \\
China
}

\maketitle

\begin{abstract}
    \noindent
Autoscaling is a key mechanism in model-as-a-service systems, where fast scale-up is
essential because the scale-up time directly determines how long a model
service remains overloaded---which in turn governs serving quality.
The state-of-the-art approach---model sleeping, adopted by mainstream serving engines including vLLM and
SGLang---achieves near-zero-cost scale-up by pausing pre-spawned engines on the
GPU.
However, sleeping is not a free lunch: each sleeping model must keep its GPU
context resident, and this cost grows linearly with the number of models,
degrading serving performance.
Because providers must support many models,
they face a dilemma between fast scale-up for all models and high
service performance.

In this paper, we show that a single server can keep many models asleep without paying
this GPU context tax.
The key observation is that the contexts of sleeping models are largely
duplicated: the driver (CUDA) context can be shared across models, and their GPU
kernel binaries overlap heavily due to common operators and libraries.
{\sys} deduplicates this context by transparently delegating each engine's GPU context allocation
to a central daemon through API remoting;
therefore, a single server can keep far more models asleep than current engines allow.
Our evaluation shows that {\sys} matches the scale-up time of vanilla vLLM sleep mode
while substantially reducing the GPU memory footprint, improving end-to-end
multi-model serving performance.
\end{abstract}


\section{Introduction}
\label{sec:intro}

\noindent
Model as a service---deploying multiple models on one or multiple shared clusters to serve LLM requests from applications---has become
an important component of modern computing infrastructure~\cite{aegaeon,openrouter,huggingface}.
Autoscaling---dynamically scaling an engine up and down on spare GPUs to serve a model---is essential for model-as-a-service systems.
The reasons are twofold. 
First, a service provider typically deploys models from many vendors (e.g., DeepSeek, Qwen, GLM, and MiniMax),
each of which offers models of different sizes.
Second, the request arrival rate of a model fluctuates over time~\cite{aegaeon,DBLP:conf/osdi/ZhangWLWS0025,DBLP:journals/corr/abs-2505-09999},
because applications choose the requested model dynamically,
depending on their budgets and capability requirements.
If we statically deploy each model, assigning it a fixed number of GPUs
without autoscaling, 
the model service becomes overloaded because it is underprovisioned when the request rate exceeds the capacity of the deployed GPUs.
Conversely, GPUs become idle because of overprovisioning when the request rate decreases.
Autoscaling dynamically adjusts each model's GPU allocation according to its changing request rate.

Efficient scale-up is a key system requirement for autoscaling.
Scale-up is triggered when the request rate increases,
but during the scale-up interval, the newly provisioned GPUs still cannot serve requests,
so the model service remains overloaded.
As a result, the scale-up time directly determines the duration of this overload,
so model-as-a-service providers
seek to minimize it.

Unfortunately, current model autoscaling has a long scale-up time because
starting an engine capable of serving a model requires a complex control plane to create the necessary GPU contexts (\textsection{\ref{sec:app:problem}}).
To reduce this delay,
the state-of-the-art method---adopted by standard engines including vLLM and SGLang---is model sleeping~\cite{aegaeon,vllm-sleep-mode}. 
To avoid creating GPU contexts during startup,
the system spawns multiple engines offline, each capable of serving one model online.
After all the engines finish initialization, they are paused on the host server.
Thus, a scale-up during online serving directly wakes up the engine, enabling near-zero-cost scale-up.

Model sleeping is not a free lunch. 
To keep one model asleep, each engine must keep a substantial amount of context on the GPU---including the driver context (CUDA context) and the model's GPU kernel binaries.
This places a heavy burden on the GPU because GPU memory is limited and precious:
in our experiment, sleeping 24 models consumes up to 27\,GB (34\%) of an 80\,GB GPU's memory,
and this occupancy scales linearly with the number of sleeping models. 
This memory footprint can further
degrade throughput by up to 54\% when serving one model,
because modern models reserve GPU memory for intermediate results (KV cache) to accelerate computation.
Sleep mode must preserve the costly parts of this context on the GPU because, unlike model parameters, which can be restored quickly~\cite{DBLP:conf/osdi/ZhangWLWS0025},
these context components cannot be restored quickly.
This poses a dilemma for the service provider: 
to enable fast scale-up for all models, the server pays online service overhead,
while to ensure high service performance,
the provider can only sleep a few models on each server,
so the service still pays the scale-up costs. 

In this paper, we show that we can increase the number of sleeping models on one server without paying a linear GPU-context memory tax.
The key observation is that the context for each sleeping model contains many duplicate components (up to 99\% similarity between two popular models, as shown in {\fig{fig:app-model-sim}}):
the driver context can be naturally shared across models,
i.e., engines created later can reuse the driver context of an earlier engine,
and the GPU kernel binaries of different models also substantially overlap because current models share operators and libraries
(\textsection{\ref{sec:phos-coldstart}}).
By deduplicating this context, we can dramatically reduce the memory required
to support more sleeping models on one server.

To realize GPU context deduplication,
we adopt a separation design that delegates each engine's GPU context allocation
to a central system daemon, where we can identify and avoid creating unnecessary contexts.
To be compatible with existing engines, we leverage API remoting~\cite{DBLP:journals/corr/abs-2401-13354,DBLP:conf/hpdc/BecchiSGPRC12} 
to implement this delegation transparently. 
One challenge of deduplication is that many internal GPU buffers allocated during 
initialization cannot be effectively deduplicated. 
Fortunately, such buffers can be created on the fly much faster than the other context components,
so we treat them differently: we leave them unallocated while the model sleeps
and reconstruct them at wake-up time.

We have implemented deduplication from scratch and integrated it with vLLM's sleep mode. 
Our evaluation shows that {\sys} achieves a scale-up time similar to that of vanilla vLLM sleep mode
while reducing the GPU memory footprint by 98.1\% when sleeping 23 models.
This saving improves end-to-end serving performance:
in a multi-model trace replay, {\sys} reduces average time to first token by
87.0\% over vanilla vLLM, which can sleep only a limited number of instances within the memory budget.
{\sys} instead sleeps all required models without incurring the GPU memory burden of vanilla sleep mode.

\paragraph{Contributions.} In summary, this paper makes the following contributions:
\begin{itemize}
\item The first characterization of the GPU-context memory tax of model sleeping and the opportunity to deduplicate it across models (\textsection{\ref{sec:app:problem}}, \textsection{\ref{sec:phos-coldstart}}).
\item {\sys}, a transparent GPU-context deduplication system with delegated context management in a service daemon, 
shares duplicate driver contexts and kernel binaries, 
and reconstructs non-deduplicable buffers only at wake-up time (\textsection{\ref{sec:phos-coldstart}}).
\item An implementation integrated with vLLM sleep mode and an evaluation showing that 
{\sys} retains its near-zero-cost scale-up while reducing GPU-context memory use and improving end-to-end multi-model serving performance (\textsection{\ref{sec:eval}}).
\end{itemize}

{\sys} is open-sourced at \blue{\url{https://github.com/blitz-serving/PhoenixOSv1}}.

\section{System Model and Problem Statement}
\label{sec:bg}
\label{sec:app:problem}

\begin{figure}[tbp]
    \centering
    \includegraphics[width=0.65\linewidth]{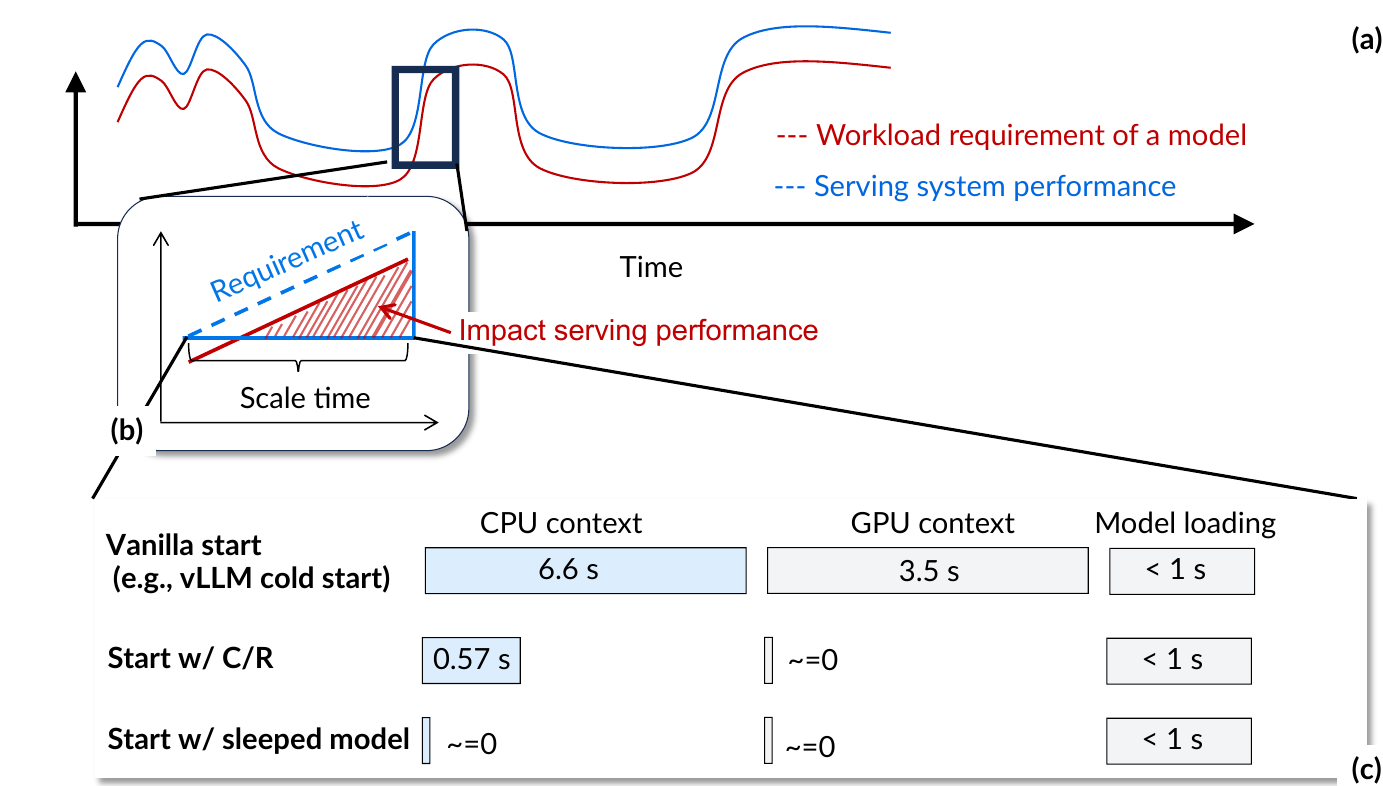}
    \caption{%
    (a) How autoscaling copes with dynamic workloads.
    (b) The impact of scale-up speed on serving performance.
    (c) A breakdown of autoscaling time across different approaches.
    }
    \label{fig:app-coldstart}
\end{figure}

\paragraph{Multi-model serving and autoscaling.}
AI service providers typically host many models of various sizes from different
vendors simultaneously in a shared GPU cluster~\cite{aegaeon,openrouter,huggingface},
because no single model dominates user demand: models differ in capability and cost.
Each model occupies one or more dedicated GPUs due to its size.

A key challenge for model providers is allocating GPU resources to each model.
Static GPU allocation---allocating
a fixed number of GPUs to each model at deployment time---is suboptimal because
per-model request rates fluctuate over time~\cite{aegaeon,DBLP:conf/osdi/ZhangWLWS0025,DBLP:journals/corr/abs-2505-09999}.
Consider assigning a fixed GPU budget to each model.
When the provisioned GPUs exceed the actual requirement,
GPUs are wasted.
On the other hand, if insufficient GPUs are provisioned, 
requests queue and experience high latency.

Modern service providers therefore use \emph{autoscaling} to adjust GPU allocation per model
in response to traffic ({\fig{fig:app-coldstart}}(a)).
Specifically, autoscaling makes
GPU provisioning for each model elastic:
when a model's load rises,
the scheduler picks one or more idle GPUs, 
starts one or more inference engines on the host servers, 
and loads the model weights to expand the model's serving capacity (scale up).
When a model's load falls, the scheduler deallocates a subset of the model's GPUs, 
releasing them back to the pool for other models
to save GPU time (scale down). 

A key system requirement of autoscaling is \emph{fast startup} during scale-up:
until the newly launched inference engines become ready to serve, 
requests to the model being scaled up must queue;
a slow startup therefore directly causes request stalls 
and SLO violations during the scaling window ({\fig{fig:app-coldstart}} (b)).
The faster the scale-up, the smaller the performance penalties.

\paragraph{Goal and design space overview.}
Our design goal is to reduce the cold start overhead of 
autoscaling a serving instance. 
To understand how to achieve this,
we first break down the cold start into two stages 
and then analyze the detailed steps involved to
determine how much we can optimize them.
\fig{fig:app-coldstart}(c) illustrates the two stages:
a \emph{control plane} that builds the serving engine's runtime context on the CPU and GPU,
and a \emph{data plane} that loads the model weights into GPU memory.

The control plane builds two contexts.
Building the CPU context starts the Python interpreter that runs
the inference engine for a model.
The interpreter
imports the serving stack (e.g., PyTorch and vLLM), 
loading tens of thousands of Python modules
and instantiating engine components such as the request scheduler.
Building the GPU context initializes the GPU driver---e.g., by
creating the CUDA primary context~\cite{cuda-primary-ctx} on the target device---
loads the model's GPU kernels into device memory,
and constructs the runtime handles the engine uses 
to launch GPU work---CUDA streams, cuBLAS, cuDNN, and so on.

The data plane copies the model weights from host memory or external
storage into GPU memory.
Its latency is bounded below by the weight size divided by the link bandwidth (e.g., PCIe or RDMA),
and has been optimized to approach the hardware limit through recent
work~\cite{DBLP:conf/osdi/ZhangWLWS0025} 
on network-based multicast weight loading.

If the server starts from a clean state (without any optimizations),
these stages could take seconds or even minutes,
as we describe in \textsection{\ref{sec:eval}}.
As a result,
prior work has proposed various methods to mitigate the slow startup issue.
In this paper, we focus on techniques to reduce the control-plane overhead,
as the data plane has been optimized to approach the hardware limit
thanks to recent work~\cite{DBLP:conf/osdi/ZhangWLWS0025,yu2025lambdascale}.

\begin{figure}[tbp]
    \centering
    \includegraphics[width=0.65\linewidth]{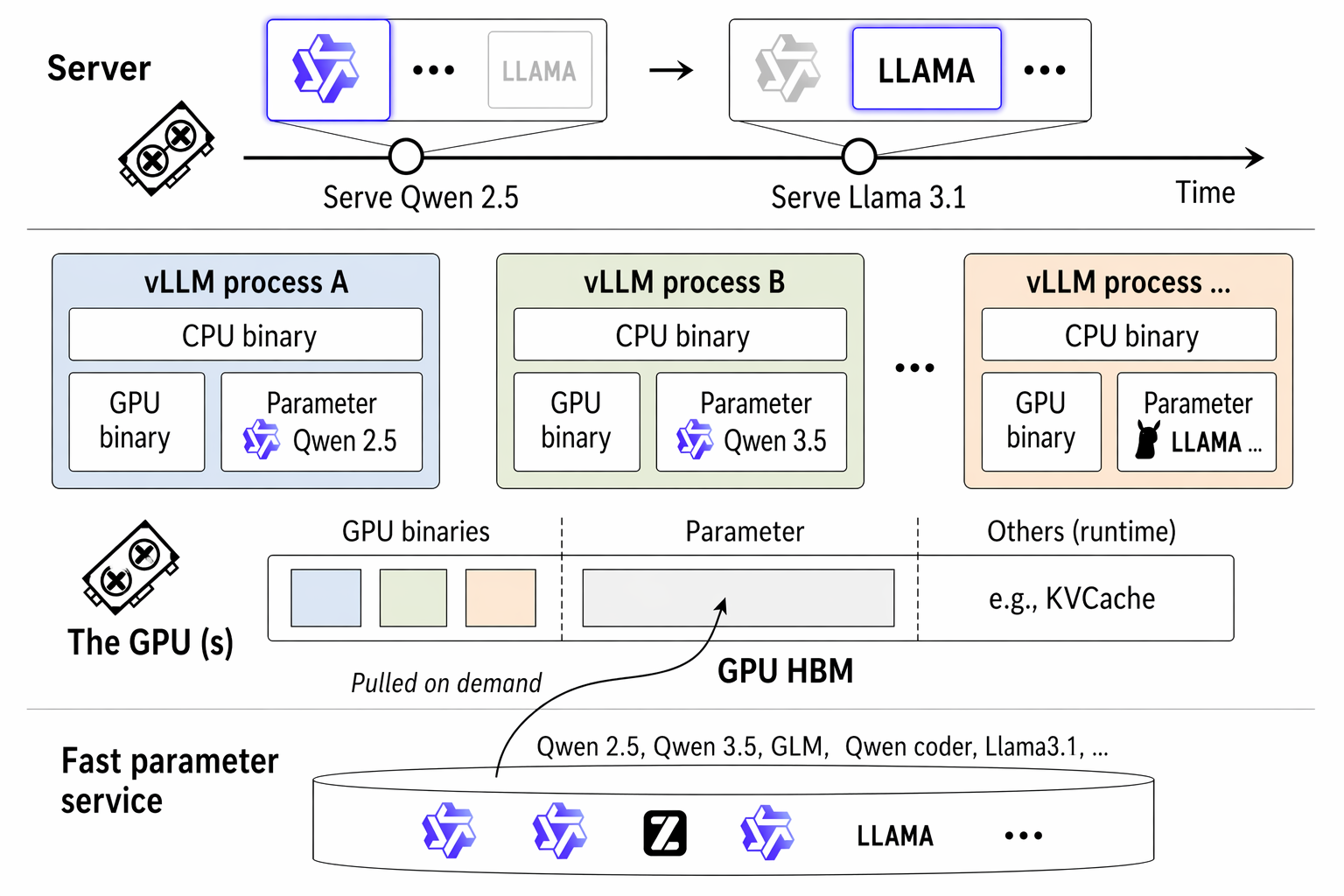}
    \caption{%
    (Top) An illustration of using model sleeping to multiplex one or more GPUs across multiple models.
    (Bottom) An illustration of the GPU memory requirement for model sleeping.
    }
    \label{fig:app-vllm-sleep}
\end{figure}

\paragraph{Current solutions: checkpoint/restore and model sleeping.}
One intuitive way to accelerate cold start is to use checkpoint/restore~\cite{DBLP:conf/sosp/WeiHSH0H0025}.
Specifically, we checkpoint an engine that has completed
control-plane initialization into an image;
on a future restart, we restore from the checkpoint image
and thus bypass all the control-plane overhead.
However, checkpoint/restore acceleration
has two issues when applied to modern inference engines.

First, performance: {\crbaseline}~\cite{DBLP:conf/sosp/WeiHSH0H0025}, the state-of-the-art system,
still incurs 0.57\,s of overhead when restoring an engine during scale-up,
because it takes time to parse the checkpoint image and restore the process's memory pages and kernel state,
as shown in {\fig{fig:app-coldstart}}(c).
In the worst case, we found that restoration can take 10--20 seconds to launch
an application because of unexplained delays in the CPU-side context restoration,
which is performed using CRIU~\cite{criu}.
Second, compatibility: modern inference engines adopt complex optimizations that
our CPU-side checkpoint/restore tool (based on CRIU~\cite{criu}) 
cannot support today, such as \texttt{io\_uring}~\cite{criu-issue-2131}.
Supporting them would require large changes to the current serving engines,
or a huge implementation effort in CRIU,
which is impractical because serving engines continue to evolve rapidly.

Another approach adopted by current systems---including 
Aegaeon~\cite{aegaeon}
and the sleep mode of vLLM and SGLang~\cite{vllm-sleep-mode}---is model sleeping.
It resembles the warm-start technique in serverless computing for cold start acceleration~\cite{DBLP:conf/usenix/ShahradFGCBCLTR20},
but applies it in the model-as-a-service setting.
Specifically, to enable fast cold start,
the sleep mode prewarms a few engines on the server, each capable of serving one model,
and then uses OS utilities~\cite{cgroup-freezer} to pause these
engines before they receive requests.
A later scale-up unpauses the required engine, bypassing
all the control-plane overhead.
After the model is scaled down, the engine is re-paused. 
In the per-GPU serving model considered here, each GPU may cache multiple sleeping engines, but at most one engine wakes up and executes on that GPU at a time.

\begin{figure}[tbp]
    \centering
    \includegraphics[width=0.6\linewidth]{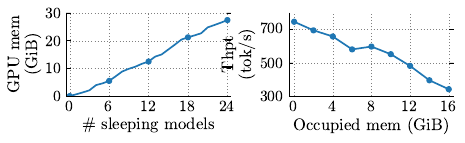}
    \caption{%
    (Left) GPU memory used to cache sleeping models grows linearly with their number.
    (Right) Serving throughput drops as available GPU memory shrinks,
    evaluated with Qwen3.6-27B on an A800 GPU\@.
    }
    \label{fig:motiv-gpu-mem}
\end{figure}

\paragraph{Key issue of model sleeping: memory scales linearly with the number of sleeping models and slows serving.}
Having discussed the issues of checkpoint/restore above,
a natural question is whether model sleeping addresses the cold-start issue.
One key problem of model sleeping is that it
trades GPU memory for fast startup.
This limits the number of models that can be prewarmed
on each machine, because the memory cost grows linearly
with the number of sleeping models.
Specifically, to make a later scale-up's control-plane cost near zero,
a paused engine keeps its GPU context resident in GPU memory and its CPU context resident in host memory.
The CPU context occupies a relatively small fraction of host memory (e.g., vLLM holds 2.9\,GB on the CPU) and
can be shared across processes using host-side shared memory,
but the GPU context is substantial and currently cannot be shared.
As a result, keeping more sleeping models on the same GPU wastes more GPU memory.
\fig{fig:motiv-gpu-mem}(left) measures this growth using vLLM v0.19.1:
caching 24 models consumes about 27\,GB of GPU memory.
Note that we have already excluded model parameters from this measurement because they can be restored quickly through
recent autoscaling techniques. 
This cost applies even when a GPU serves only one model at a time:
as long as it keeps other models sleeping, their GPU contexts stay resident.
{\sys} focuses on reducing this resident GPU-context memory cost; concurrent sharing and scheduling of multiple active engines on the same GPU are outside its scope.

This memory cost creates a trade-off for providers.
Modern LLM providers serve many models~\cite{aegaeon,openrouter,huggingface}.
Model sleeping removes the control-plane cost only for models already sleeping,
so avoiding that cost across many models requires keeping many of them asleep.
One may ask whether the memory cost of keeping many models asleep is affordable.
We found it is not, because
the resident memory hurts serving: the more models kept asleep,
the more serving performance drops.
This is because serving a model requires
storing its KV cache---the intermediate results of computation---in GPU memory.
With less GPU memory available for the KV cache, requests queue and slow down significantly~\cite{DBLP:conf/eurosys/ChengLWCC26},
and the model's context length is restricted. 
\fig{fig:motiv-gpu-mem}(right) quantifies this:
serving throughput falls from 747 to 341\,tokens/s (a 54\% drop) as the memory occupied by sleeping models grows from 0 to 16\,GB\@.
Providers therefore keep only a few models asleep today
and pay the full control-plane cost for models not kept asleep.

\section{Design of {\sys}} 
\label{sec:phos-coldstart}

\noindent
{\sys} therefore has two goals:
(1) remove the control-plane cost, even when a GPU is likely to sleep many models at the same time;
(2) avoid the GPU memory cost of keeping many models asleep.

\begin{figure}[tbp]
    \centering
    \includegraphics[width=\linewidth]{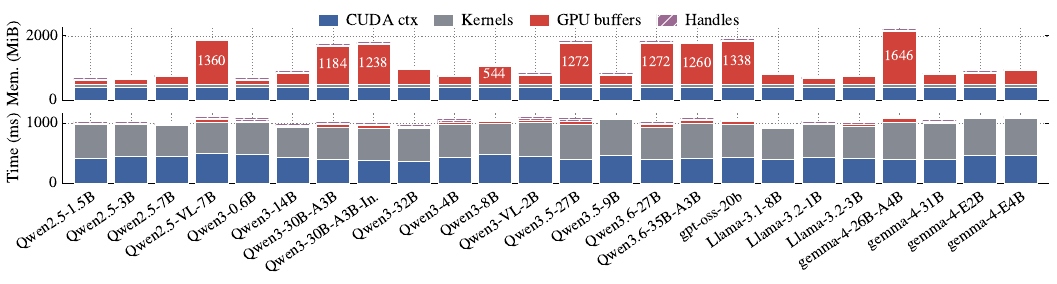}
    \caption{
        (Top) GPU memory used (without parameters) by each sleeping model, broken down by resource type,
        and (Bottom) the time required to restore this memory.
    }
    \label{fig:eval-ctx-breakdown}
\end{figure}

\paragraph{Key design opportunity: GPU context across sleeping models is redundant.}
{\sys} builds upon model sleeping to remove the control-plane cost,
which enables fast scale-up.
To further address its memory expansion issues,
we make a key observation:
their GPU contexts contain substantial duplicate content.
If we can deduplicate this content,
we can dramatically reduce the GPU memory required to keep multiple models asleep.

More specifically, we first
break down the GPU memory used for each sleeping model in detail
in {\fig{fig:eval-ctx-breakdown}} to assess how much of
their context we can deduplicate.
Four components contribute to GPU-resident memory:
CUDA context---the GPU runtime environment containing 
control state for virtual-memory management and synchronization events~\cite{cuda-programming};
GPU kernels---the compiled GPU code that runs the model's inference;
GPU buffers---memory allocated for the model's internal use;
and handles---registries for resources and libraries 
used to access the GPU, 
such as CUDA streams and cuBLAS~\cite{cuBLAS}.

\begin{figure}[tbp]
    \centering
    \includegraphics[width=0.6\linewidth]{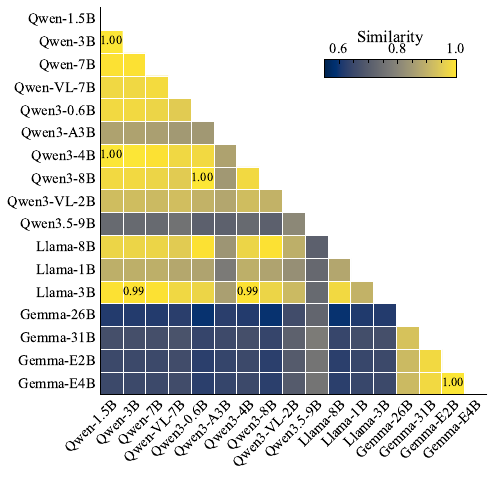}
    \caption{%
    Pairwise similarity of GPU kernels across popular models. 
    }
    \label{fig:app-model-sim}
\end{figure}

Except for the GPU buffers, which are model-dependent,
and the handles, whose memory usage is negligible,
we found that most content can be deduplicated.
First, 
the CUDA context can be shared across models. 
This is straightforward because the driver context itself is model-independent,
i.e., the context used by one model can trivially be used for another.
Moreover, the kernels used by models also have a high degree of overlap:
{\fig{fig:app-model-sim}} provides the pairwise similarity of GPU kernels across models
excluding GPU buffers.
We choose models with high download counts on Hugging Face~\cite{huggingface}
because they represent popular models typically served by model-as-a-service providers.
Most model pairs have 59--99\% kernel similarity.
Despite differences in architecture and size,
existing models share similar kernels because 
(1) many operations, such as matrix multiplication, are common across models,
and (2) a narrow set of libraries (e.g., FlashInfer~\cite{flashinfer}) provides these operations
because writing them efficiently for GPUs is non-trivial.
As a result, deduplicating CUDA context and kernels can dramatically
reduce the GPU memory used to keep multiple models asleep.

One may notice that the GPU buffers still account for much of the GPU memory used
for some models (e.g., Qwen3.6-27B, as shown in the top panel of {\fig{fig:eval-ctx-breakdown}}).
Nevertheless, we found that such buffers can be dynamically restored quickly:
as shown in the bottom panel of {\fig{fig:eval-ctx-breakdown}},
creating them from scratch only takes $\sim$30\,ms,
compared with the $\sim$430\,ms required to create a CUDA context.
Such buffers are typically reserved for internal 
use at runtime,
so we can safely avoid storing them on the GPU
and create them on demand.

\begin{figure}[tbp]
    \centering
    \includegraphics[width=0.65\linewidth]{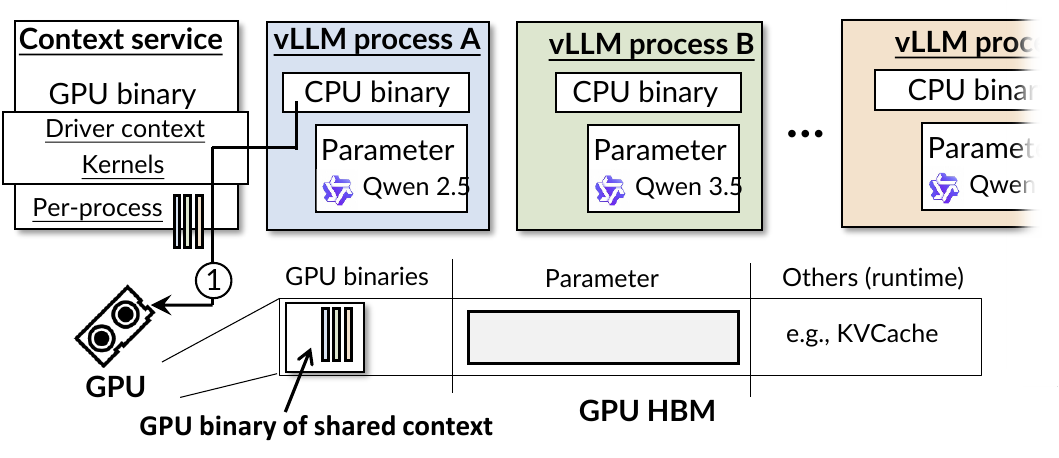}
    \caption{%
    Architecture of {\sys}'s context service, which shares GPU context across
    sleeping models.
    }
    \label{fig:app-arch}
\end{figure}

\paragraph{System architecture.}
To realize deduplication,
we need to ensure that all the GPU contexts from
different sleeping models are managed together by a central service,
so that it can detect duplicate components and avoid
creating duplicate resources.
Thus, {\sys} deploys a per-machine system daemon for managing GPU contexts 
from different models,
as shown in {\fig{fig:app-arch}}. 
Specifically, unlike the existing sleep mode, in which each model's engine process
manages its own GPU context (the top panel of {\fig{fig:app-vllm-sleep}}),
the context service manages GPU kernel binaries
in a single CUDA context shared by all models assigned to a GPU,
along with per-model resources, such as customized kernels.
On a server with multiple GPUs, the daemon manages one such context
per GPU, allowing wake-ups on different GPUs to proceed independently.
The kernels are properly deduplicated:
before a model loads a kernel binary,
we check whether the load is necessary,
i.e., whether the same binary has already been loaded.
If so, we skip the load to reduce the GPU memory footprint.

Below, we describe {\sys} in detail:
how we construct the context service offline
and how the online serving system uses it to support
inference requests after an engine wakes up.

\paragraph{Offline phase: sleep the models and initialize the context service.}
The offline phase initializes the sleeping engines to support efficient online scale-up.
This phase is optional but recommended:
otherwise, the system initializes models in the context service on demand,
i.e., upon first startup,
incurring a one-time scale-up delay.
In practice, this process can also be performed incrementally,
by reusing scale-up instances or starting models during request idle time,
thereby amortizing its cost and avoiding subsequent scale-up delays.

The offline phase works as follows. 
Given $N$ models that need fast startup via sleep mode on one server,
we first initialize a context service as a background daemon 
that manages the GPU resources.
After the daemon is initialized, 
we start the engines (e.g., vLLM/SGLang) one by one, 
putting each engine into sleep mode once it finishes starting up.
Once asleep, an engine's per-process state is paused, 
while its GPU context is held by the context service rather than by the engine itself.
To achieve this,
the engine does not create GPU resources directly; instead, it
delegates resource creation to the context service.
We make this delegation transparent to the engine using a well-established GPU virtualization technique: API remoting~\cite{DBLP:journals/corr/abs-2401-13354,DBLP:conf/hpdc/BecchiSGPRC12}.
Specifically, we use \texttt{LD\_PRELOAD} to hook all the GPU resource creation calls
at the CUDA Driver API boundary and execute them in the context service,
requiring no source-code changes to inference engines.

After receiving a resource creation request,
the context service examines whether the resource already exists.
If the resource (e.g., a kernel) has not been created, 
the context service creates it and records the creation information,
so that later requests can recognize it.
We use a lookup table that encodes the API semantics:
for example, 
\texttt{cuModuleLoadData} loads a kernel binary into the GPU\@.
A subsequent call with the same binary will be recognized as a duplicate,
so the service simply skips the creation.
The table incorporates all CUDA APIs
that inference engines use for GPU resource management
and therefore covers all GPU-context allocations.

GPU buffers require special handling.
If an engine requests a buffer allocation,
the service creates a virtual placeholder for it but delays the real allocation until the online phase. 
This is fine because, as shown in the bottom panel of {\fig{fig:eval-ctx-breakdown}} and discussed above,
creating these resources online is fast.
The GPU buffer's contents are checkpointed in host memory to ensure
correct restoration.

\paragraph{Online phase: supporting model wake-up and inference requests.}
During the online phase, after an engine wakes up to serve requests,
we only need to direct it to use the GPU context stored by the context service for inference.
We use the same resource-management mechanism described above:
API remoting.
When an engine invokes a GPU kernel,
the call is first redirected to the context service,
which creates the required GPU buffers if necessary and then invokes the kernel. 
Remoting the GPU API may add performance overhead to inference.
This overhead is negligible in the evaluated workloads
(\textsection{\ref{sec:eval-e2e}}) thanks to optimizations including batching
and synchronization, 
because inference is primarily bottlenecked by GPU execution
rather than CPU-side kernel launches~\cite{DBLP:journals/corr/abs-2401-13354}.

If the model's context is not already held by the context service,
we perform the sleep procedure described above to store its context there,
so that future wake-ups avoid a cold start.
Storing a context that the service does not yet hold typically reserves
additional GPU memory to accommodate the new model.
Left unchecked, this could lead to a large memory footprint
as we keep more models asleep.
To address the issue,
we set a maximum amount of GPU memory that may be reserved for sleeping models,
and evict lesser-used models from the context service if necessary.
Specifically, 
we use a least-frequently-used (LFU) strategy to evict sleeping models.
Whether other strategies can further improve system performance
is beyond the scope of our work.

\section{Evaluation}
\label{sec:eval}
\label{sec:eval-setup}

\noindent
We have implemented {\sys} in 10,825 lines of Rust. Our evaluation answers three
questions:
(1) Does {\sys} improve end-to-end multi-model serving performance under
autoscaling?
(2) Does it remove the linear GPU-memory cost of sleeping many models?
(3) Does it preserve the low scale-up latency of vanilla sleep mode?

\paragraph{Testbed.}
We evaluate {\sys} on a single server with eight NVIDIA A800 GPUs, each with
80\,GB of HBM\@.
All measurements reported in this section use a single GPU.
The server has two Intel Xeon Platinum 8350C CPUs providing 64 cores in total and
1\,TB of DRAM\@.
The server runs CUDA 12.9~\cite{cuda123} and vLLM v0.19.1.

\paragraph{Baselines.}
We compare {\sys} with three baselines.
\emph{Cold start} launches a vanilla vLLM inference engine from scratch when a model is
requested.
\emph{Vanilla sleep} uses vLLM sleep mode: it discards the KV cache, keeps model
parameters in host memory, and later wakes the same vLLM engine.
{\crbaseline} uses checkpoint and restore to start the engine:
after sleep, the process is checkpointed; during scale-up, the process is
restored and then woken up.

\paragraph{Workloads and metrics.}
We describe the end-to-end workload in \textsection{\ref{sec:eval-e2e}}.
Unless stated otherwise, we use 16 Qwen and Qwen-VL models with high download counts
on Hugging Face~\cite{huggingface}.

We report GPU memory footprint, startup latency, end-to-end time to first token
(TTFT), TTFT without queueing delay, and input- and output-token throughput. End-to-end
TTFT spans the interval from when a request is issued to when its first token is produced and
may therefore include queueing behind earlier requests. TTFT without queueing delay
spans the interval from when the system begins scheduling the target model to the first token; if
the model is inactive, it includes its cold-start or wake-up time.

\paragraph{Measurement methodology.}
We use NVML~\cite{nvidia2026nvml} to measure GPU memory currently occupied by the vLLM process.
When reporting GPU-context memory, we measure GPU memory after putting the vLLM engine
to sleep; consequently, model weights are excluded.

\subsection{End-to-end experiments}
\label{sec:eval-e2e}

\begin{figure}[tbp]
    \centering
    \includegraphics[width=\linewidth]{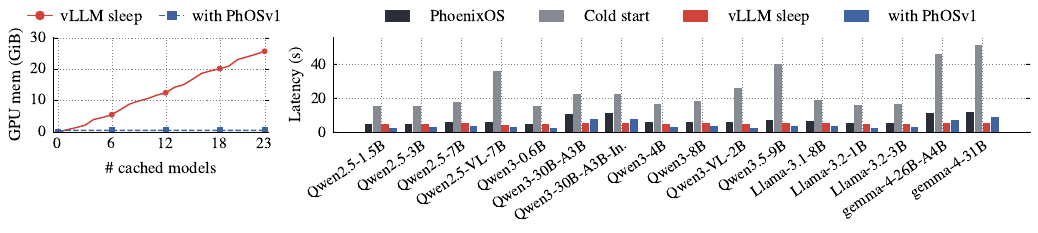}
    \caption{
        (Left) GPU memory used to cache sleeping
        models under vanilla vLLM sleep mode and {\sys}.
        (Right) Startup
        latency across models for cold start, vanilla sleep, {\crbaseline},
        and {\sys}.
        Weight-loading times for cold start and {\crbaseline} are normalized to remove storage I/O overhead.
    }
    \label{fig:eval-startup-combined}
\end{figure}

\begin{figure}[tbp]
    \centering
    \includegraphics[width=0.6\linewidth]{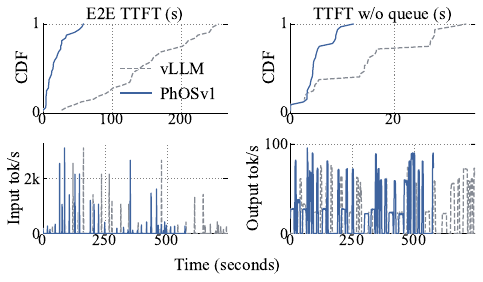}
    \caption{
        End-to-end serving performance.
    }
    \label{fig:eval-startup-timeline}
\end{figure}

\paragraph{Evaluation setup.}
To evaluate whether {\sys} improves user-visible multi-model serving performance
under autoscaling,
we construct a multi-model end-to-end trace using Aegaeon's trace
methodology~\cite{aegaeon}.
Specifically,
we serve 10 popular Qwen models ranging from 0.6B to
32B parameters~\cite{huggingface}.
The 10-minute ShareGPT~\cite{sharegpt} trace is generated with Poisson arrivals,
with each model receiving requests at a rate of 0.3 requests/min.
We set this per-model arrival rate to represent scenarios with
infrequent requests to many long-tail models, where model sleeping is required for fast startup~\cite{aegaeon}.
The vanilla-sleep baseline uses a 6\,GB sleep-cache budget, reserving the remaining
GPU memory for the active model's weights and KV cache; it keeps as many sleeping
vLLM instances as possible under this budget, puts a model back to sleep after
inference, evicts models using LFU when the budget is exceeded, and cold-starts a model
when it is not cached.
This 6\,GB budget is the largest that
preserves vLLM's default GPU-memory allocation; a larger budget would reduce the
KV-cache capacity and thus the supported context length.
We use the same vLLM configuration for both {\sys} and vanilla sleep.

\paragraph{Overall performance.}
Figure~\ref{fig:eval-startup-timeline} shows that {\sys} reduces the average
end-to-end TTFT from 146.82\,s to 19.09\,s, an 87.0\% improvement over vanilla
vLLM instances limited by the sleep-cache budget.
It also reduces p50 TTFT from 142.39\,s to 14.65\,s, while finishing the same
workload in a shorter active serving window (578\,s vs. 728\,s).
The improvement comes from caching and sleeping the required models without
consuming the large linear GPU-context memory of vanilla sleep mode; the baseline
can only keep a limited set of models warm under the memory budget, so requests
for uncached models pay cold-start delay and create queueing that dominates
end-to-end TTFT.

\paragraph{Overhead of our technique.}
To evaluate the inference overhead introduced by our shared context,
we report the throughput (tokens per second) of {\sys} and vanilla vLLM in
the bottom panels of Figure~\ref{fig:eval-startup-timeline}.
We can see that {\sys}'s throughput
is at least 97.3\% of vanilla vLLM's,
indicating that the overhead introduced by the shared context is small.

\subsection{GPU context memory reduction and startup performance}
\label{sec:eval-factor}

\noindent
Figure~\ref{fig:eval-startup-combined} (left) evaluates whether
{\sys} removes the GPU-memory growth that prevents vanilla sleep mode from
scaling to many cached models.
We use 23 models from Figure~\ref{fig:eval-ctx-breakdown}; model weights are
excluded.
Figure~\ref{fig:eval-startup-combined} shows that vanilla sleep consumes
25.67\,GiB of GPU memory after caching 23 models, whereas {\sys}+sleep uses only
0.49\,GiB, reducing the memory footprint by 98.1\%.
The savings come from sharing the CUDA context and deduplicating common kernel
binaries in the context service instead of letting each sleeping engine hold a
private GPU context.

Figure~\ref{fig:eval-startup-combined} (right) further evaluates whether
{\sys} preserves the low wake-up latency of sleep mode while adding a context
service connection and API-remoting layer.
We use 16 models and compare cold start, waking up a sleeping vLLM
model, {\crbaseline}, and {\sys}.
Figure~\ref{fig:eval-startup-combined} shows that {\sys} keeps startup
latency comparable to vanilla sleep, averaging 4.12\,s across 16 models, compared
with 4.89\,s for vanilla sleep and 24.41\,s for cold start.
Startup remains low because the expensive control-plane context is already
available in the service, and the per-model cost of connecting to the context service is
small, averaging 72\,ms.

\section{Discussion}
\label{sec:limitations}

\noindent
This section discusses several assumptions and limitations of {\sys}.
First, {\sys} assumes that engines sharing a CUDA context are trusted.
This assumption applies to model-as-a-service deployments~\cite{aegaeon},
where the service provider manages the engines that
serve different models.
Sharing a context across untrusted engines is outside the scope of our work.

Second, {\sys} uses API remoting to implement context sharing,
which may introduce inference overhead.
This implementation choice allows {\sys} to share a context without modifying
NVIDIA's closed-source CUDA driver.
Implementing context sharing directly in the driver could avoid the
remoting overhead.

Finally, {\sys} does not support CUDA graphs.
Checkpointing and restoring these graphs requires preserving or reconstructing
the device pointers and kernel references embedded in graph nodes, as
discussed in prior work~\cite{DBLP:conf/asplos/ZengXGCL25,DBLP:journals/corr/abs-2604-06664}.
We leave support for CUDA graphs to future work.

\section{Related Work}
\label{sec:related}

\paragraph{Fast startup for serverless and LLM serving.}
Prior work accelerates serverless container startup through caching,
checkpoint/restore, and remote fork, reusing initialized execution
environments~\cite{sock,DBLP:conf/asplos/DuYXZYQWC20,DBLP:conf/osdi/WeiLW0Y0023}.
Related techniques accelerate VM startup~\cite{DBLP:conf/eurosys/AoPV22}
and GPU checkpoint/restore~\cite{DBLP:conf/sosp/WeiHSH0H0025,DBLP:conf/fast/ZengRSL26}.
For LLM serving, ServerlessLLM exploits storage locality for fast parameter
loading~\cite{DBLP:journals/corr/abs-2401-14351}.
BlitzScale combines network transfer and cooperative execution for fast
autoscaling~\cite{DBLP:conf/osdi/ZhangWLWS0025}, and FaaSwap swaps weights under
SLO constraints~\cite{DBLP:journals/corr/abs-2306-03622}.
Medusa and Foundry accelerate CUDA graph setup through offline
materialization~\cite{DBLP:conf/asplos/ZengXGCL25,DBLP:journals/corr/abs-2604-06664}.
These approaches reduce parameter-loading costs or reconstruct selected execution
state, complementing {\sys}'s deduplication of resident GPU context across
sleeping engines.
Model sleeping in vLLM/SGLang and Aegaeon~\cite{vllm-sleep-mode,aegaeon}
avoids startup overhead but retains a separate GPU context for each model
(\textsection{\ref{sec:app:problem}}).

\paragraph{OS primitives for GPUs.}
NVIDIA MPS supports concurrent CUDA processes through shared scheduling
resources~\cite{nvidia-mps}. {\sys} instead targets GPU-context deduplication
across sleeping engines for fast scale-up.
MIG partitions a GPU into isolated compute and memory instances~\cite{nvidia-mig},
whereas we target serving workloads that use a full GPU for the active model.

TGS shares GPUs across containers through rate control and unified-memory
oversubscription~\cite{tgs}. REEF provides microsecond-scale kernel preemption;
Paella offers software-defined GPU scheduling; and XSched extends preemptive
scheduling to diverse XPUs~\cite{reef,DBLP:conf/sosp/NgD023,DBLP:conf/osdi/ShenHL0025}.
GPU API remoting and virtualization decouple applications from physical
devices~\cite{DBLP:journals/corr/abs-2401-13354,DBLP:conf/hpdc/BecchiSGPRC12}.
{\sys} builds on API remoting to provide GPU-context deduplication across
sleeping models.

\section{Conclusion}
\label{sec:concl}

\noindent
We contribute GPU-context deduplication, the first technique that lets a single
server keep many models asleep for fast autoscaling without paying a linear GPU memory
tax.
The key insight is that the GPU contexts of sleeping models contain largely
duplicate components that can be shared across models through a central daemon,
transparently to existing inference engines.
Our evaluation shows that {\sys} preserves the fast scale-up of model sleeping while
substantially reducing its memory footprint, improving end-to-end multi-model
serving performance.

\bibliographystyle{unsrt}
\bibliography{pos}

\end{document}